\documentclass[useAMS,usenatbib]{mnras}

\usepackage{graphics}
\usepackage{amssymb}
\usepackage{amsfonts}
\usepackage{wasysym}
\usepackage{epsfig}
\usepackage{breakurl}

\usepackage{longtable}
\usepackage{times}
\usepackage{mathptmx}
\usepackage[usenames,dvipsnames]{color}

\title[The auroral radio emission of HD142301]
{The polarization mode of the auroral radio emission from the early-type star HD142301}
\author[P. Leto et al.]
{P. Leto$^{1}$ \thanks{E-mail: paolo.leto@inaf.it},
C. Trigilio$^{1}$,
L. M. Oskinova$^{2,3}$,
R. Ignace$^{4}$,
C. S. Buemi$^{1}$,
G. Umana$^{1}$,
\newauthor F. Cavallaro$^{1}$,
A. Ingallinera$^{1}$,
F. Bufano$^{1}$,
N. M. Phillips$^{5,6}$,
C. Agliozzo$^{5}$,
\newauthor L. Cerrigone$^{6}$,
H. Todt$^{2}$,
S. Riggi$^{1}$,
F. Leone$^{7}$
\\
$^{1}$INAF - Osservatorio Astrofisico di Catania, Via S. Sofia 78, 95123 Catania, Italy\\
$^2$Institute for Physics and Astronomy, University Potsdam, 14476 Potsdam, Germany\\
$^3$Kazan Federal University, Kremlevskaya Str 18, Kazan, Russia\\
$^4$Department of Physics \& Astronomy, East Tennessee State University, Johnson City, TN 37614, USA\\
$^5$European Southern Observatory, Alonso de C\'{o}rdova 3107, Vitacura, Santiago, Chile\\
$^6$Joint ALMA Observatory, Alonso de C\'{o}rdova 3107, Vitacura, Santiago, Chile\\
$^7$Universit\'{a} degli studi di Catania, Via S.Sofia 78, I-95123 Catania, Italy
}
\begin{document}

\date{}

\pagerange{\pageref{firstpage}--\pageref{lastpage}} \pubyear{}

\maketitle

\label{firstpage}

\begin{abstract}
We report the detection of the auroral radio emission from the early-type magnetic star HD\,142301.
New VLA observations of HD\,142301 detected highly polarized amplified emission occurring at fixed stellar orientations.
The coherent emission mechanism responsible for the stellar auroral radio emission amplifies the radiation within a narrow beam, 
making the star where this phenomenon occurs similar to a radio lighthouse.
The elementary emission process responsible for the auroral radiation
mainly amplifies one of the two magneto-ionic modes of the electromagnetic wave.
This explains why the auroral pulses are highly circularly polarized.
The auroral radio emission of HD\,142301 is characterized by a reversal
of the sense of polarization as the star rotates.
The effective magnetic field curve of HD\,142301 is also available making it possible to correlate
the transition from the left to the right-hand circular polarization sense
(and vice-versa) of the auroral pulses with the known orientation of the stellar magnetic field.
The results presented in this letter have implications for the estimation of the dominant magneto-ionic mode amplified within the HD\,142301 magnetosphere.
\end{abstract}



\section{Introduction}

The auroral radio emission is a kind of coherent radiation detected from
magnetized planets of the solar system \citep{zarka98}, some brown dwarfs \citep*{hallinan_etal15,kao_etal16,pineda_etal17}, 
and three early-type magnetic stars: CU\,Vir \citep{trigilio_etal00,trigilio_etal11},
HD\,133880 \citep{das_etal18}, and HD\,142990 \citep{lenc_etal18}. 
The amplification mechanism is the Electron Cyclotron Maser (ECM) powered by an unstable energy distribution 
(see \citealp{treumann06} for an extensive review of the ECM),
that the electrons moving within a density-depleted magnetospheric cavity can develop,
e.g. the loss-cone \citep{wu_lee79,melrose_dulk82} or the horseshoe 
\citep{winglee_pritchett86} distributions.
The ECM mainly amplifies one of the two magneto-ionic modes, each with opposite circular polarization sense.
This explains the high polarization degree of the stellar auroral radio emission.

In the case of the ECM driven by the loss-cone unstable energy distribution,
the dominant magneto-ionic mode depends on
the physical conditions of the regions where the ECM operates,
especially the local magnetic field strength ($B$) and  plasma density ($N_{\mathrm e}$).
In particular, the growth rate of the extraordinary (X) mode is faster when $\nu_{\mathrm p} / \nu_{\mathrm B} \leq 0.3$--0.35
(where $\nu_{\mathrm p}= 9 \times 10^{-6} \sqrt{N_{\mathrm e}}$ GHz is the plasma frequency and
$\nu_{\mathrm B} =2.8 \times 10^{-3} B/{\mathrm G}$ GHz is the gyro-frequency),
whereas the ordinary (O) mode is favored in 
the range 0.3--$0.35 < \nu_{\mathrm p} / \nu_{\mathrm B} \leq 1$ \citep*{sharma_vlahos84,melrose_etal84}.
The polarization sense of the X-mode
is in accordance with the helicity of the electrons moving in the magnetic field.
In the case of a magnetic field vector with a positive component along the line of sight 
the electrons are seen rotate counter clockwise (CCW); but clockwise (CW) if the magnetic field vector is oppositely oriented. 
The corresponding polarization sense will be RCP or LCP,
in accordance with the IAU and IEEE orientation/sign convention. 
The O-mode has instead the opposite sense of polarization with
respect to the helicity of the electrons. 
For the horseshoe driven ECM  the waves are purely amplified in the X-mode,
as directly measured in the case of Auroral Kilometric Radiation (AKR) from the Earth \citep{ergun_etal00}.

In this letter we report the radio light curves of HD\,142301 at two frequencies (1.5 and 5.5 GHz).
The 5.5 GHz emission is compatible with non-thermal incoherent gyro-synchrotron,
that is commonly observed in many early-type magnetic stars \citep*{drake_etal87,linsky_etal92,leone_etal94}.
The flux of HD\,142301 at 5.5 GHz is rotationally modulated, 
similar to other early-type magnetic stars
\citep{leone91,leone_umana93,leto_etal06,leto_etal12,leto_etal17a,leto_etal18}.
This is in accordance with the radio emission 
from an optically thick Rigidly Rotating Magnetosphere (RRM) \citep{trigilio_etal04,townsend_owocki05}
shaped like an oblique dipole \citep{babcock49,stibbs50}.
The 1.5 GHz light curve is instead characterized by the presence 
of intense strongly polarized emission, occurring at well defined stellar rotational phases.
This is a clear signature of auroral radio emission from HD\,142301.
Furthermore, the sense of the circular polarization of the auroral radiation reverses as the star rotates.

HD\,142301 is the fourth early-type magnetic star where the auroral radio emission has been detected
and the first showing reversal of the sense of
circular polarization of the auroral pulses. 
The helicity reversal of the auroral pulses
has already been observed at the bottom of the main sequence in three Ultra Cool Dwarfs  
\citep*{hallinan_etal07,lynch_etal15,kao_etal16}.
The UCDs are fast rotators ($P_{\mathrm {rot}}$ few hours) and faint sources. 
Measurements of their average magnetic fields (levels of kG) were reported \citep{reiners_basri10},
but, at this time, effective magnetic field curves 
of these dwarf stars are unavailable,
where the effective magnetic field ($B_{\mathrm e}$) is defined as the average over the whole visible disc of 
the magnetic field vector components along the line of sight.
The effective magnetic field curve of HD\,142301 is instead well known \citep*{landstreet_etal79}.
Hence, this early-type magnetic star offers the unique possibility 
to correlate the circular polarization sense of its auroral radio emission
with the magnetic field vector orientation.

\section{The magnetic field geometry of HD\,142301}
\label{orm}

HD\,142301 (3 Sco) is a He-weak \citep{leone_lanzafame97} B8III/IV type star located in the Sco-Cen stellar association.
The measured stellar luminosity and effective temperature (listed in Table~\ref{par_star}) 
were used to estimate the radius and the stellar gravity \citep{kochukov_bagnulo06}:
 $R_{\ast}=2.52$ R$_{\odot}$; $\log g=4.3$.
The rotation period of HD\,142301 is: $P_{\mathrm {rot}} \approx 1.46$ days \citep{shore_etal04}.
Adopting the projected rotational velocity $v \sin i =74.8$ km s$^{-1}$ \citep{glebocki_gnacinski05}  
and the stellar radius given above, 
by using the relation $v \sin i = 2 \pi R_{\ast} \sin i/ P_\mathrm{rot}$
we estimate an inclination of the stellar rotation axis $i=59^{\circ}$.

In the framework of the Oblique Rotator Model (ORM) the misalignment between magnetic and rotation axes (angle $\beta$)
can be derived by using the relation: $\tan \beta \tan i = (1-r)/(1+r)$ \citep{preston67},
where $r$ is the ratio between the minimum and maximum effective magnetic field ($B_{\mathrm e}$).
For HD\,142301 $r=-0.41$ \citep{landstreet_etal79}, it follows that $\beta=55^{\circ}$.

For the case of a magnetic field topology  described by a simple dipole,
the polar magnetic field strength ($B_{\mathrm p}$) can be derived by using the relation \citep{schwarzschild50}:
\begin{equation}
\label{eq_bpolo}
B_{\mathrm p}= |B_{\mathrm e}(max)|  \frac{4(15-5u)}{15+u} \cos (i-\beta)
\end{equation}
where $|B_{\mathrm e}(max)|=4100$ G \citep{landstreet_etal79} and $u$ is the limb darkening coefficient.
The effective magnetic field measurements of HD\,142301 were performed analyzing 
the H$\beta$ line wings \citep{landstreet_etal79}, and the corresponding average limb darkening coefficient is $u=0.36$
\citep{claret_etal11}.
Consequently, the polar field strength of HD\,142301, estimated by using the Eq.~\ref{eq_bpolo}, is $B_{\mathrm p}=14100$ G.
However, a field topology more complex than a simple dipole
has been proposed to describe the magnetic field of HD\,142301 \citep{glagolevskij10}.

\section{Observations and data reduction}
\label{sec_vla}

In this letter we report on observations of HD\,142301 performed by the 
Karl G. Jansky Very Large Array (VLA), operated by the
National Radio Astronomy Observatory
NRAO), at the L and C bands (CODE: 16A-043).
HD\,142301 was observed in different epochs distributed between February and August 2016,
when the VLA array configurations were C, CnB, and B,
by using the 8-bit sampler.
For the L-band measurements, 
the above observing setup allows to cover a frequency range of 1 GHz width
that is centered at 1.5 GHz.
The C-band measurements provide a bandwidth of 2 GHz centered at 5.5 GHz. 

The total (Stokes~$I$) and the circularly polarized (Stokes~$V$) intensities are obtained combining
the two opposite polarized components of the electromagnetic wave (RCP and LCP):
$I= ({\mathrm{RCP}}+{\mathrm{LCP}})/2$; $V=({\mathrm{RCP}}-{\mathrm{LCP}})/2$. 
HD\,142301 was always observed  
close to the primary beam center, then,
to analyze the circular polarization state of HD\,142301
only the standard calibration is required.
The flux density scale was calibrated observing the standard flux calibrator 3C286.
The complex gain was determined by observing J1522-2730.

\begin{table}
\caption[ ]{HD\,142301  stellar parameters \citep{kochukov_bagnulo06}}
\label{par_star}
\footnotesize
\begin{tabular}[]{lcc}
\hline
Parameter                                                      & Symbol                         &                                            \\                 
\hline
Mass  [M$_{\odot}$]                                 &{$M_{\ast}$}                & {$4.3\pm 0.2$}                                          \\
Luminosity  [L$_{\odot}$]                                 &{$L_{\ast}$}                & {$10^{2.53 \pm 0.15} $}                                          \\
Effective Temperature [K]                          &$T_\mathrm{eff}$          & $15600\pm400$                                   \\
\hline
\end{tabular}
\end{table}

The VLA measurements of HD\,142301 were edited and calibrated by using the standard calibration pipeline,
working on the Common Astronomy Software Applications ({\sc casa}).
Sub-bands contaminated by strong RFI were flagged.
Images at the sky position of HD\,142301 were performed by using the task {\sc clean}.
The source was not resolved by the array spatial resolution. 
To measure the flux density of HD\,142301, a gaussian bi-dimensional fit of the radio source at the sky position of HD\,142301
was performed in the cleaned maps, both for the Stokes~$I$ and $V$.
The flux density uncertainty was estimated summing in quadrature the map noise with the error related
to the gaussian fitting process.
With a scan duration of about 2 and 7 minutes long, respectively for the C and L bands,
the map noise level was $\approx 0.03$ mJy/beam for both bands.
The average uncertainties of the flux density measurements 
for each observing band are: $\approx 5 \%$ at 1.5 GHz;
$\approx 3 \%$ at 5.5 GHz.

The measurements reported in this letter have been complemented by further old VLA observations of HD\,142301 performed at similar frequencies,
published \citep*{linsky_etal92,leone_etal94,leone_etal96} and un-published (our projects: AL346 and AL388).
Raw data were directly retrieved from the VLA archive,
then calibrated and imaged by using the standard calibration tasks in the {\sc casa} package.

\begin{figure}
\resizebox{\hsize}{!}{\includegraphics{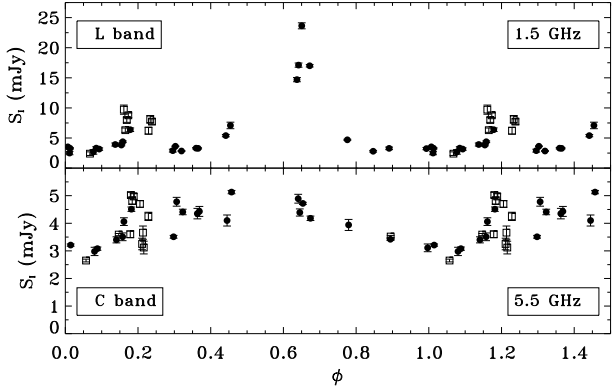}}
\caption{Total intensity (Stokes~$I$) radio light curves of  HD\,142301.
Top panel: 1.5 GHz measurements.
Bottom panel: 5.5 GHz measurements. 
The filled circles represent the new wide band radio measurements reported in this paper, the open squares refer to
other radio measurements.} 
\label{fig_data}
\end{figure}

\section{The radio light curves of HD\,142301}
\label{sec_aurora}

\begin{figure}
\resizebox{\hsize}{!}{\includegraphics{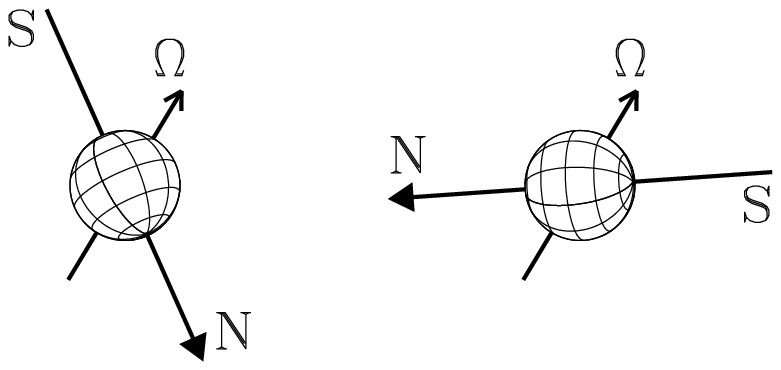}}
\resizebox{\hsize}{!}{\includegraphics{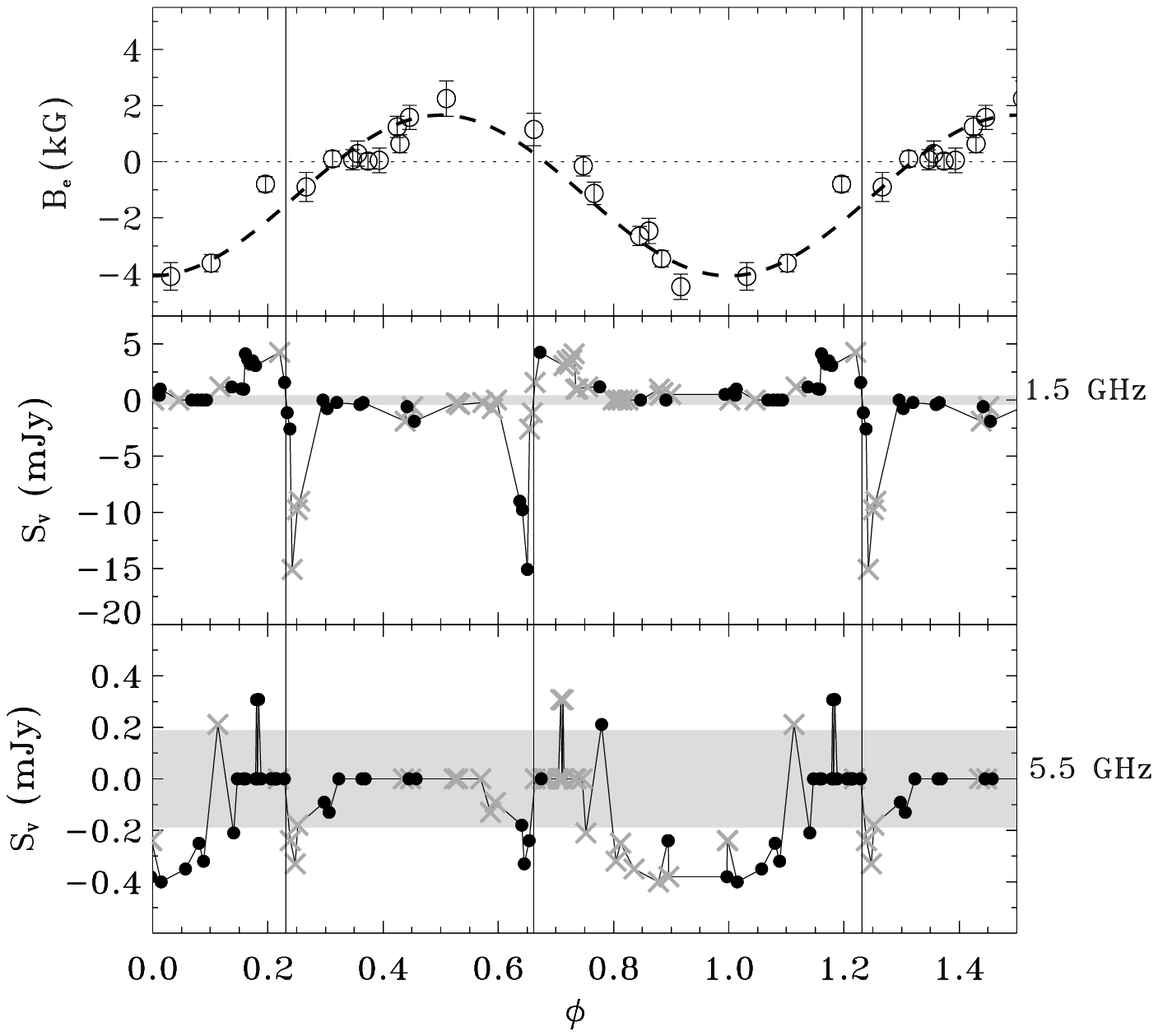}}
\caption{Top panel: effective magnetic field measurements of HD\,142301 (open circles) taken by \citet{landstreet_etal79}.
The dashed line represents the magnetic curve calculated by using the ORM geometry of HD\,142301.
Cartoons showing the stellar orientations corresponding to the two magnetic curve extrema 
are also pictured (observer on the right).
Other panels: light curves of the circularly polarized emission (Stokes~$V$) at 1.5 and 5.5 GHz.
The filled dots represent the radio measurements. 
The vertical continuous lines locate the rotational phases where the circular polarization helicity reversal occurs.
The radio emission behavior is symmetric respect to the intermediate stellar orientation between these two phases. 
The mirrored radio measurements are displayed by the gray `{\sc x}' symbol.
The light grey bands around the zero level highlights the average $3\sigma$ threshold.}
\label{fig_b_v}
\end{figure}

The L and C bands radio measurements of HD\,142301 
were phase folded using the ephemeris given by \citet{shore_etal04}:
\begin{equation}
\label{effemeridi}
{{\mathrm {HJD}}=2~449~545.378(4)+1.45957(5)E ~{\mathrm{(days)}}}
\end{equation}
The zero-point of the stellar phases is associated with the negative magnetic extremum.
The phase folded 
radio light curves for the Stokes~$I$ are displayed in Fig.~\ref{fig_data}.
The light curves for the Stokes~$V$ are displayed in the middle (1.5 GHz) 
and bottom (5.5 GHz) panels of Fig.~\ref{fig_b_v}. 
The effective magnetic field measurements \citep{landstreet_etal79} phased using
Eq.~\ref{effemeridi} are displayed
in the top panel of Fig.~\ref{fig_b_v},
the theoretical magnetic curve
calculated by using the ORM geometry given in Sec.~\ref{orm} 
(using the procedure of \citealp{leto_etal16}) is also pictured.

The C-band measurements show radio emission from HD\,142301 that
is clearly rotationally modulated,
both for the total and the circularly polarized emission.
In particular, the fraction of the circularly polarized 5.5 GHz radiation
reaches its maximum negative value ($\approx -10\%$) when
the south magnetic pole of HD\,142301 is seen at small viewing angle
(bottom panel of Fig.~\ref{fig_b_v}), in agreement with
the magnetic field vectors 
mainly radially oriented towards the stellar surface.
For the ORM geometry of HD\,142301 (sec.~\ref{orm}) the north magnetic  pole is not seen at small viewing angle 
(see the cartoons pictured in Fig.~\ref{fig_b_v}).
This explains why the right hand circularly polarized emission from the northern hemisphere is not detected at 5.5 GHz.
This is the usual behavior 
for such stars \citep{trigilio_etal04,leto_etal06,leto_etal12,leto_etal17a,leto_etal18},
that is explained in the framework
of the magnetically confined 
wind shock (MCWS) model  \citep{babel_montmerle97}. 
The wind plasma streams, from opposite hemispheres,
move along the dipolar magnetic field lines. At the magnetic equator 
the streams collide and shock, radiating thermal X-rays. 
Far from the star, the magnetic fields no longer dominate the ionized trapped matter.
In the resulting current sheets, electrons can be accelerated up to relativistic energies.
These non-thermal electrons move within a magnetospheric cavity
and radiate at radio wavelengths by the incoherent gyro-synchrotron emission mechanism.

The L-band  total intensity of HD\,142301 is rotationally modulated too,
in addition it is
characterized by amplified emission occurring 
at certain stellar rotational phases
(see top panel of Fig.~\ref{fig_data}).
These radio peaks are highly circularly polarized (see middle panel of Fig.~\ref{fig_b_v}).
The 1.5 GHz circularly polarized emission strongly depends on the stellar orientation.
This is a signature of radio emission confined within a narrow beam,
producing radio pulses like a radio lighthouse.
Like the case of CU\,Vir \citep{trigilio_etal11},
the lighthouse effect of HD\,142301 can be explained as auroral radio emission
tangentially beamed along the auroral cavity walls \citep{louarn_lequeau96}.
The auroral cavity coincides with the magnetospheric cavity where the non-thermal electrons,
responsible for the incoherent gyro-synchrotron emission, freely propagate.
Such non-thermal electrons can develop an un-stable energy distribution
(loss-cone or horseshoe)
that drives the ECM coherent emission mechanism responsible for the stellar auroral radio emission. 
Reversal in the sense of the circular polarization within the auroral pulses of
HD\,142301 has been clearly detected.
The two phases ($\phi=0.23$; $\phi=0.66$), where the polarization reversal occurs, 
are highlighted by the vertical lines in Fig.~\ref{fig_b_v}.

The dominant magneto-ionic mode of
the stellar auroral radio emission can be right or left-handed circularly polarized, depending on the magnetic field vector orientation 
in the region where the coherent emission arises.
Then, the measurement of a reversal of the sense of the circular polarization 
localizes the stellar orientation where the magnetic axis is perpendicular to the line of sight.
Follow that, the plane of symmetry of HD\,142301 (the plane containing the rotation and magnetic axes)
crosses the line of sight at $\phi_{\mathrm {sim}}=0.445$ and $\phi_{\mathrm {sim}}+0.5$.
Due to the low declination ($\delta \approx -25^{\circ}$) of  HD\,142301, this is visible for about 6 hours above $20^{\circ}$
of the VLA horizon, that is a short fraction of the stellar rotation period ($P_{\mathrm {rot}} \approx 35$ hours).
The light curve of the HD\,142301 auroral radio emission has been then performed collecting measurements
randomly distributed over the time.
To compensate the not optimal phase sampling, we exploit the symmetric behavior of the radio emission from the HD\,142301 magnetosphere.
The Stokes~$V$ measurements have been mirrored with respect  to $\phi_{\mathrm {sim}}$,
the so produced light curves are
pictured in the middle and bottom panels of Fig.~\ref{fig_b_v}.

\begin{figure}
\resizebox{\hsize}{!}{\includegraphics{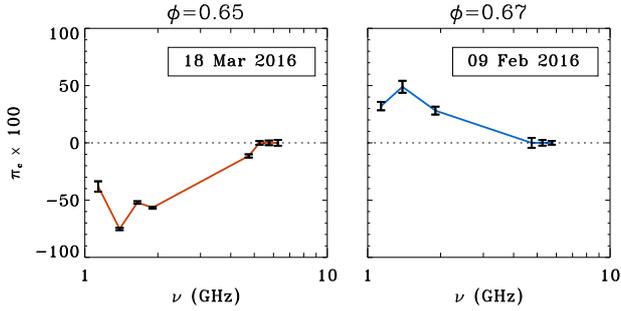}}
\caption{Spectral dependence of the  fraction of the circularly polarized emission of HD\,142301 performed at the phases just
preceding (left panel) and following (right panel) the stellar orientation where a circular polarization reversal occurs.}
\label{spe}
\end{figure}

The L-band observations, tracking the circular polarization helicity reversal
at $\phi=0.23$, are narrow band VLA measurements, performed on 06 Aug 1993, 
whose average total intensity was already published \citep{leone_etal96}. On the other hand,
the four measurements close to $\phi=0.66$ are the new wide-band VLA data, that we acquired at different epochs:
09 Feb 2016; 18 Mar 2016; 27 Jun 2016; 16 Jul 2016.
These wide-band radio measurements can be used 
to probe the spectral dependence of the auroral radio emission of HD\,142301.
The fractional circular polarization 
($\pi_{\mathrm c}={\mathrm {Stokes}~V}/{\mathrm {Stokes}~I}$)
spectra performed at  $\phi=0.65$ and  $\phi=0.67$
are shown in Fig.~\ref{spe}.
The analyzed epochs (09 Feb 2016 and 18 Mar 2016) have the nearest rotation phases
showing auroral emission of opposite helicity.
The circular polarization fraction is maximum at $\nu \approx 1.4$ GHz.
At $\phi=0.65$ we measured  $\pi_{\mathrm c} \approx -75$\%, whereas at  $\phi=0.67$ $\pi_{\mathrm c} \approx +50$\%.

\section{Discussion and conclusions}
\label{sec_concl}

The ECM emission mechanism amplifies the radiation at frequencies close 
to the first few harmonics of the local gyro-frequency ($\nu_{\mathrm B}$),
that is proportional to the local field strength ($B$).
On the basis of the well known radial dependence of a simple magnetic dipole
($B = B_{\mathrm p} r^{-3}$, where $r$ is the radial distance), 
the 1.5 GHz auroral radio emission of HD\,142301 ($B_{\mathrm p}=14100$ G) arises from 
a magnetospheric layer at $r \approx 3$ R$_{\ast}$, corresponding to rings at $\approx 2$ R$_{\ast}$ above the magnetic poles.
The 5.5 GHz ECM emission, if detectable, would be generated
$\approx 1$ R$_{\ast}$ above the stellar surface.

As discussed in the previous section, 
the behavior of the HD\,142301 emission at 5.5 GHz
is canonical incoherent gyro-synchrotron emission from a dipole shaped stellar magnetosphere.
While, the behavior of HD\,142301 measured at 1.5 GHz  cannot be explained as canonical gyro-synchrotron emission.
The 1.5 GHz emission is strongly dominated by the coherent auroral radio emission.
The main observing features of the 
auroral radio emission from HD\,142301,
phase location of the pulses and circular polarization sense transition, 
have been simulated by using the three-dimensional model to study the visibility of the stellar auroral pulses radiated
from a thin auroral cavity \citep{leto_etal16}.
This model has wide application. It can be used to simulate the auroral radio emission from a
dipole shaped stellar magnetosphere in general. In fact, this model was also applied to study
the auroral pulses from the UCD TVLM\,513-46546 \citep{leto_etal17b}.

\begin{figure}
\resizebox{\hsize}{!}{\includegraphics{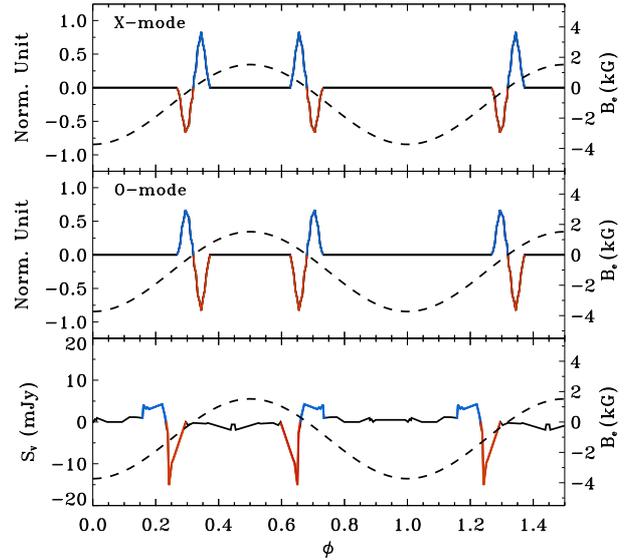}}
\caption{ 
Top and middle panels, simulated HD\,142301 auroral radio emission for the X and O-mode.
Bottom panel, observed light curve of the HD\,142301 auroral emission.
The effective magnetic field curve (dashed line),
performed by using the ORM geometry of HD\,142301, 
is superimposed to the simulated and observed light curves of the auroral emission.
}
\label{sim}
\end{figure}

In the framework of the adopted model, the beam pattern of the ECM amplified radiation is assumed strongly anisotropic.
The auroral emission is radiated in direction almost perpendicular to the local magnetic field vector and
constrained in the plane tangent to the auroral ring (within a narrow beaming angle).
Due to refraction effects, the ray path of the auroral radiation is upward deviated.

To simulate the features of the HD\,142301 auroral emission, the ORM geometry outlined in Sec.~\ref{orm} was adopted.
The model free parameters able to control the auroral beam pattern were widely varied. 
The fine tuning of these parameters is beyond the scope of this paper;
we report only that, when the model parameters make the simulated auroral emission from HD\,142301 detectable,
the main features of the simulated auroral pulses remain unchanged.
In Fig.~\ref{sim} two representative simulated auroral light curves are displayed.
Top panel of Fig.~\ref{sim} refers to RCP polarized auroral emission 
from the northern hemisphere and LCP from the southern, 
in this case the favored magneto-ionic mode amplified by the ECM mechanism is the X-mode.
The middle panel refers to the opposite case, in accord with the O-mode auroral emission.
For comparison, in the bottom panel of Fig.~\ref{sim} the observed HD\,142301 auroral emission is pictured again.

Looking at Fig.~\ref{sim} it is clear that the sequence of the circular polarization helicity inversion
is in accord with the observations
when the left-handed circularly polarized emission arises from the north hemisphere and, vice-versa,
the right-hand from the southern.
The behavior of the HD\,142301 radio emission observed at 1.5 GHz
is compatible with O-mode auroral emission.
This has implications on the unstable electron distribution driving the ECM
(loss-cone versus horseshoe).
In fact, a pure horseshoe-driven ECM is able to produce only X-mode auroral emission \citep{ergun_etal00},
not compatible with the observed helicity of the HD\,142301 pulses.
Whereas, the loss-cone driven ECM is able to amplify also the O-mode \citep{lee_etal13}.

The above conclusion allows us to constrain the physical conditions of the magnetospheric regions
where the ECM emission mechanism operates.
Using the relationship between the plasma and the gyro-frequency that favors the O-mode amplification \citep{sharma_vlahos84,melrose_etal84},
we estimate a plasma density $N_{\mathrm e} \approx 10^{9}$--$10^{10}$ cm$^{-3}$ 
in the regions where the ECM amplification mechanism takes place.
The presence of circumstellar plasma located in the deep magnetospheric layers of HD\,142301
has been reported \citep{shore_etal04}, similar to the case of the hot magnetic stars with a rigidly rotating magnetosphere
\citep{groote_hunger97,townsend_etal05}.

The absence of C-band auroral emission from HD\,142301 
highlights that the coherent emission has a high frequency cutoff at $\nu \approx 5$ GHz (see Fig.~\ref{spe}).
This could be an indication of the inability of the high-frequency auroral radiation to escape from the magnetosphere of HD\,142301.
In fact, the electromagnetic waves freely propagate only if the plasma refraction index is real.
In the case of wave propagation parallel to magnetic field,
the refraction index for the O-mode is 
$n^2_{\mathrm{refr}}  = 1-\nu^2_{\mathrm{p}} / (\nu^2 + \nu\nu_{\mathrm B})$ 
\citep{klein_trotter84}.
Following the above relation, if $N_{\mathrm e} > 5 \times 10^{11}$  cm$^{-3}$ the refraction index is
not compatible with the 5 GHz wave propagation.
The 5 GHz auroral emission originates $\approx 1$ R$_{\ast}$ above the HD\,142301 surface, then,
the possible presence of such high-density plasma will inhibit its propagation.
Further, high density plasma close to the surface prevents formation of a density-depleted cavity,
necessary to develop an unstable electron energy distribution to drive the ECM.
This could cause the 5 GHz amplification mechanism to fade.
It is worth to note, that clouds of high density plasma ($N_{\mathrm e} > 10^{11}$--$10^{12}$  cm$^{-3}$)
have been observed within the RRM of some early B type stars \citep{grunhut_etal12,rivinius_etal13}.
The 1.5 GHz auroral radiation arises from farther regions, 
with densities favorable for the O-mode amplification 
($N_{\mathrm e} \approx 10^{9}$--$10^{10}$ cm$^{-3}$).
At these lower densities in the magnetosphere, 
the index of refraction is instead compatible with wave propagation at 1.5 GHz.

The comparison between the observed and the simulated auroral pulses
allows us also to confirm the suspected non-dipolar nature of the magnetic field topology of HD\,142301 
as suggested by \citet{glagolevskij10}.
For a dipole shaped ORM, the auroral pulses coincide with the phases where the magnetic dipole axis is located in the plane of the sky
(null effective magnetic field).
Looking at the bottom panel of Fig.~\ref{sim}, 
it is clear that the phases where the helicity polarization reversal occurs
are close, but not coinciding, with the phases where
the effective magnetic field is null. 
Furthermore the minimum phase interval between the simulated peaks is $\Delta \phi = 0.36$,
versus a longer measured phase interval ($\Delta \phi = 0.43$). 
This highlights that the ORM magnetic field geometry, 
deduced by the measurements of the effective magnetic field at the stellar surface,
is only as a first approximation to the true field topology that dominates the magnetosphere of HD\,142301.

\section*{Acknowledgments}
We are grateful to the referee for the very constructive comments which
helped us to improve the paper. The 
National Radio Astronomy Observatory is a facility of the National Science 
Foundation operated under cooperative agreement by Associated Universities, Inc.
This work has extensively used  the NASA's Astrophysics Data System, and the 
SIMBAD database, operated at CDS, Strasbourg, France. 
LMO acknowledges support by the DLR grant 50\,OR\,1508.



\end{document}